\documentstyle[prd,preprint,tighten,aps,floats,eqsecnum,%
amssymb,amsfonts,newlfont,graphicx]{revtex}
\newcommand{\midx}[1]{{#1}}
\begin{document}
\draft
\preprint{
\begin{tabular}{r}
   DFTT 55/99
\\ hep-ph/9912211
\end{tabular}
}
\title{Four-Neutrino Mixing and Long-Baseline Experiments}
\author{Carlo Giunti}
\address{INFN, Sezione di Torino, and Dipartimento di Fisica Teorica,
Universit\`a di Torino,\\
Via P. Giuria 1, I--10125 Torino, Italy}
\maketitle
\begin{abstract}
We consider the two four-neutrino schemes
that are compatible with
all neutrino oscillation data.
We present the range of the corresponding mixing parameters
allowed by
the results of neutrino oscillation experiments.
We discuss the implications for long-baseline experiments
and we suggest the possibility to reveal the presence of
CP-violating phases
through a comparison of the oscillation probability
measured in long-baseline experiments
with the corresponding average oscillation probability
measured in short-baseline experiments.
\end{abstract}


\section{Introduction}
\label{Introduction}

The three indications in favor of neutrino oscillations
found in solar neutrino experiments \cite{exp-sun},
in atmospheric neutrino experiments \cite{exp-atm} and
in the accelerator LSND experiment \cite{LSND}
require the existence of three independent scales
of neutrino mass-squared differences
(see \cite{Giunti-70-99+Giunti-nufac-99}):
$\Delta{m}^2_{\mathrm{sun}} \sim 10^{-6} - 10^{-4} \, \mathrm{eV}^2$
(MSW)
or
$\Delta{m}^2_{\mathrm{sun}} \sim 10^{-11} - 10^{-10} \, \mathrm{eV}^2$
(VO)
\cite{analysis-sun},
$\Delta{m}^2_{\mathrm{atm}} \sim 10^{-3} - 10^{-2} \, \mathrm{eV}^2$
\cite{SK-lp99,analysis-atm},
$\Delta{m}^2_{\mathrm{SBL}} \sim 1 \, \mathrm{eV}^2$
\cite{LSND}.
Hence,
at least four massive neutrinos must exist in nature.

Considering the minimal possibility
of four massive neutrinos
\cite{four-models,four-phenomenology,%
BGG-AB-96,Barger-variations-98,BGGS-AB-99,%
BGG-bounds-98,BGG-CP-98,DGKK-99},
we have the mixing relation
\begin{equation}
\nu_{\alpha L} = \sum_{k=1}^4 U_{\alpha k} \, \nu_{kL}
\qquad
(\alpha=e,\mu,\tau,s)
\,,
\label{mixing}
\end{equation}
that connects the flavor neutrino fields
$\nu_{{\alpha}L}$
($\alpha=e,\mu,\tau,s$)
with the fields $\nu_{kL}$
($k=1,\ldots,4$)
of neutrinos with masses $m_k$.
$U$ is the $4{\times}4$ unitary mixing matrix.
The field $\nu_{sL}$
describes sterile neutrinos,
that do not participate to weak interactions
and do not contribute to the invisible decay width of the $Z$-boson,
whose measurement have shown that the number of light active
neutrino flavors is three
(see \cite{PDG-98}),
corresponding to $\nu_e$, $\nu_\mu$ and $\nu_\tau$.

The purpose of this paper is to present the
implications of the results of neutrino oscillation experiments
for the mixing of four light massive neutrinos.
We will discuss also some implications
for long-baseline neutrino oscillation experiments
\cite{LBL}.

The plan of the paper is as follows.
In Section \ref{The mixing schemes}
we introduce the two four-neutrino mixing
schemes that are allowed by the data
and we review some of their properties.
In Section \ref{Constraints from oscillation experiments}
we derive the constraints on four-neutrino mixing
obtained from the results of neutrino oscillation experiments.
In Sections
\ref{Long-baseline disappearance experiments}
and
\ref{Long-baseline appearance experiments}
we discuss some implications
for long-baseline disappearance
and appearance experiments, respectively.
In Section \ref{Conclusions} we summarize our achievements.

In the discussion of long-baseline experiments
we will neglect matter effects,
that may be relevant for long-baseline
accelerator experiments whose beams propagate
for hundreds of kilometers in the crust of the Earth,
but are rather complicated to study in the framework of
four-neutrino mixing (see \cite{DGKK-99})
and will be treated in detail in another article
\cite{DGKK-preparation}.

\section{The mixing schemes}
\label{The mixing schemes}

It has been proved in \cite{BGG-AB-96,Barger-variations-98,BGGS-AB-99}
that among all the possible four-neutrino mixing schemes only two
can accommodate the results of all neutrino oscillation experiments:
\begin{equation}
\mbox{(A)}
\qquad
\underbrace{
\overbrace{m_1 < m_2}^{\mathrm{atm}}
<
\overbrace{m_3 < m_4}^{\mathrm{sun}}
}_{\mathrm{SBL}}
\,,
\qquad
\mbox{(B)}
\qquad
\underbrace{
\overbrace{m_1 < m_2}^{\mathrm{sun}}
<
\overbrace{m_3 < m_4}^{\mathrm{atm}}
}_{\mathrm{SBL}}
\,.
\label{AB}
\end{equation}
These two spectra are characterized by the presence of two pairs
of close masses separated by a gap of about 1 eV
which provides the mass-squared difference\footnote{
In the following we will denote the mass-squared differences
$m_k^2 - m_j^2$
with
$\Delta{m}^2_{kj}$.
}
$ \Delta{m}^2_{\mathrm{SBL}} = \Delta{m}^2_{41} $
responsible for the
short-baseline (SBL) oscillations observed in the LSND experiment.
In the scheme A,
we have
$ \Delta{m}^2_{\mathrm{atm}} = \Delta{m}^2_{21} $
and
$ \Delta{m}^2_{\mathrm{sun}} = \Delta{m}^2_{43} $,
whereas in scheme B,
$ \Delta{m}^2_{\mathrm{atm}} = \Delta{m}^2_{43} $
and
$ \Delta{m}^2_{\mathrm{sun}} = \Delta{m}^2_{21} $.
In order to treat simultaneously the two schemes
A and B,
let us rename the mass eigenstates indices as
\begin{eqnarray}
&&
\midx{c} = 1
\,,\,
\midx{d} = 2
\,,\,
\midx{a} = 3
\,,\,
\midx{b} = 4
\,,\,
\mbox{in scheme A}
\,,
\label{AB01}
\\
&&
\midx{a} = 1
\,,\,
\midx{b} = 2
\,,\,
\midx{c} = 3
\,,\,
\midx{d} = 4
\,,\,
\mbox{in scheme B}
\,.
\label{AB02}
\end{eqnarray}
With this convention,
in both schemes
$\nu_{\midx{a}}$
and
$\nu_{\midx{b}}$
are the two massive neutrinos whose mass-squared difference
is responsible for solar neutrino oscillations
and
$\nu_{\midx{c}}$
and
$\nu_{\midx{d}}$
are the two massive neutrinos whose mass-squared difference
generates atmospheric neutrino oscillations:
\begin{equation}
\Delta{m}^2_{\mathrm{sun}} = \Delta{m}^2_{\midx{b}\midx{a}}
\,,
\qquad
\Delta{m}^2_{\mathrm{atm}} = \Delta{m}^2_{\midx{d}\midx{c}}
\,.
\label{AB03}
\end{equation}

For
$\Delta{m}^2_{\mathrm{SBL}}$
in the following we will use the range
\begin{equation}
0.20 \, \mathrm{eV}^2
\lesssim
\Delta{m}^2_{\mathrm{SBL}}
\lesssim
2.0 \, \mathrm{eV}^2
\,,
\label{LSND-range}
\end{equation}
that is allowed by the 99\% CL
maximum likelihood analysis of the LSND data
in terms of two-neutrino oscillations
\cite{LSND}
and by the 90\% CL
exclusion curves of the Bugey reactor $\bar\nu_e$
disappearance experiment \cite{Bugey-95}
and of the
BNL E776 \cite{BNL-E776} and KARMEN \cite{karmen-moriond-99}
experiments (see also \cite{Eitel-99})\footnote{
In order to obtain reliable results,
in this paper we use 99\% CL limits when possible.
Limits with 90\% CL are used only when
they are the only available information
from the corresponding experiment.
}.

The mixing in both schemes A and B
is constrained by the results of neutrino oscillation experiments
\cite{BGG-AB-96,Barger-variations-98,BGGS-AB-99,Giunti-70-99+Giunti-nufac-99}.
In order to quantify these constraints,
let us define the quantities $\Sigma_\alpha$,
with $\alpha=e,\mu,\tau,s$,
as
\begin{equation}
\Sigma_\alpha
\equiv
\sum_{k=\midx{c},\midx{d}}
|U_{{\alpha}k}|^2
\,.
\label{Sigma}
\end{equation}
Physically $\Sigma_\alpha$ quantify the mixing of the flavor neutrino $\nu_\alpha$
with the two massive neutrinos whose
mass-squared difference
is relevant for the oscillations of atmospheric neutrinos
($\nu_1$, $\nu_2$ in scheme A
and
$\nu_3$, $\nu_4$ in scheme B).

Let us write down
the expressions,
that we will need in the following,
for the oscillation probabilities
in short baseline experiments
and
in atmospheric and long-baseline experiments.
The neutrino oscillation probabilities in short-baseline (SBL)
experiments are given by \cite{BGG-AB-96}
\begin{eqnarray}
&&
P^{\mathrm{SBL}}_{\stackrel{\makebox[0pt][l]
{$\hskip-3pt\scriptscriptstyle(-)$}}{\nu_{\alpha}}
\to\stackrel{\makebox[0pt][l]
{$\hskip-3pt\scriptscriptstyle(-)$}}{\nu_{\beta}}}
=
\frac{1}{2}
\,
A_{\alpha\beta}
\left( 1 - \cos\frac{ \Delta{m}^{2}_{\mathrm{SBL}} L }{ 2 E } \right)
\qquad
(\beta\neq\alpha)
\,,
\label{Pab-SBL}
\\
&&
P^{\mathrm{SBL}}_{\stackrel{\makebox[0pt][l]
{$\hskip-3pt\scriptscriptstyle(-)$}}{\nu_{\alpha}}
\to\stackrel{\makebox[0pt][l]
{$\hskip-3pt\scriptscriptstyle(-)$}}{\nu_{\alpha}}}
=
1
-
\frac{1}{2}
\,
B_{\alpha\alpha}
\left( 1 - \cos\frac{ \Delta{m}^{2}_{\mathrm{SBL}} L }{ 2 E } \right)
\,,
\label{Paa-SBL}
\end{eqnarray}
where
$L$ is the source-detector distance,
$E$ is the neutrino energy
and the oscillation amplitudes are given by
\begin{equation}
A_{\alpha\beta}
=
4 \Bigg|
\sum_{ \scriptstyle \renewcommand{\arraystretch}{0.5}
\begin{array}{c} \scriptstyle
k=\midx{a},\midx{b}
\\ \scriptstyle
\mathrm{or}
\\ \scriptstyle
k=\midx{c},\midx{d}
\end{array}
}
U_{{\beta}k} U_{{\alpha}k}^{*}
\Bigg|^2
\,,
\qquad
B_{\alpha\alpha}
=
4 \Sigma_\alpha ( 1 - \Sigma_\alpha )
\,.
\label{SBL-amp}
\end{equation}
The two expressions for
$A_{\alpha\beta}$
obtained by summing over
$k=\midx{a},\midx{b}$
or
$k=\midx{c},\midx{d}$
are equivalent because of the unitarity of the mixing matrix.
The oscillation probabilities of neutrinos and antineutrinos
in short-baseline experiments are equal.
Since the expressions (\ref{Pab-SBL}) and (\ref{Paa-SBL})
have the same form as the usual probabilities in the case of two-neutrino
mixing,
the oscillation amplitudes
$A_{\alpha\beta}$ and $B_{\alpha\alpha}$
are equivalent to the usual parameter
$\sin^2 2 \vartheta$
used in the analysis of the experimental results
($\vartheta$ is the two-neutrino mixing angle).
The average probabilities measured by short-baseline
experiments that are sensitive to values of
$\Delta{m}^{2}$
much smaller than the actual value of
$\Delta{m}^{2}_{\mathrm{SBL}}$
are given by
\begin{equation}
\langle
P^{\mathrm{SBL}}_{\stackrel{\makebox[0pt][l]
{$\hskip-3pt\scriptscriptstyle(-)$}}{\nu_{\alpha}}
\to\stackrel{\makebox[0pt][l]
{$\hskip-3pt\scriptscriptstyle(-)$}}{\nu_{\beta}}}
\rangle
=
\frac{1}{2}
\,
A_{\alpha\beta}
\quad
(\alpha\neq\beta)
\,,
\qquad
\langle
P^{\mathrm{SBL}}_{\stackrel{\makebox[0pt][l]
{$\hskip-3pt\scriptscriptstyle(-)$}}{\nu_{\alpha}}
\to\stackrel{\makebox[0pt][l]
{$\hskip-3pt\scriptscriptstyle(-)$}}{\nu_{\alpha}}}
\rangle
=
1
-
\frac{1}{2}
\,
B_{\alpha\alpha}
\,.
\label{LBL08}
\end{equation}

The probability of
$\nu_\alpha\to\nu_\beta$
transitions in atmospheric and long-baseline experiments
is given by\footnote{
When in the following we will need to write the explicit
expressions for oscillation probabilities that are different for neutrinos
and antineutrinos,
we will consider,
for simplicity,
only neutrinos.
The corresponding expressions
for antineutrinos are obtained by taking the complex conjugate
of the elements of the mixing matrix.
}
\cite{BGG-AB-96}
\begin{equation}
P^{\mathrm{LBL}}_{\nu_\alpha\to\nu_\beta}
=
\left|
\sum_{k=\midx{a},\midx{b}}
U_{{\beta}k}
\,
U_{{\alpha}k}^{*}
\right|^2
+
\left|
U_{\alpha\midx{c}}^{*}
U_{\beta\midx{c}}
+
U_{\alpha\midx{d}}^{*}
U_{\beta\midx{d}}
\exp\!\left(
- i
\frac{ \Delta{m}^{2}_{\mathrm{atm}} L }{ 2 E }
\right)
\right|^2
\,.
\label{LBL04}
\end{equation}
The average oscillation probability
measured in experiments sensitive to
values of $\Delta{m}^2$
smaller than the actual value of $\Delta{m}^2_{\mathrm{atm}}$
is given by
\begin{equation}
\langle
P^{\mathrm{LBL}}_{\stackrel{\makebox[0pt][l]
{$\hskip-3pt\scriptscriptstyle(-)$}}{\nu_{\alpha}}
\to\stackrel{\makebox[0pt][l]
{$\hskip-3pt\scriptscriptstyle(-)$}}{\nu_{\beta}}}
\rangle
=
\left|
\sum_{k=\midx{a},\midx{b}}
U_{{\alpha}k}^{*}
U_{{\beta}k}
\right|^2
+
\left(
\sum_{k=\midx{c},\midx{d}}
|U_{{\alpha}k}|^2
|U_{{\beta}k}|^2
\right)
\,,
\label{LBL05}
\end{equation}
for both neutrinos and antineutrinos.

\section{Constraints from oscillation experiments}
\label{Constraints from oscillation experiments}

The negative results
of short-baseline
$\bar\nu_e$ and $\nu_\mu$ disappearance experiments
and the results of solar neutrino experiments
imply that
$\Sigma_e$ and $\Sigma_\mu$ are constrained by
\cite{BGG-AB-96,BGGS-AB-99}
\begin{equation}
\Sigma_e \leq a^0_e(\Delta{m}^2_{\mathrm{SBL}})
\qquad \mbox{and} \qquad
\Sigma_\mu \leq a^0_\mu(\Delta{m}^2_{\mathrm{SBL}})
\quad \mbox{or} \quad
1 - \Sigma_\mu \leq a^0_\mu(\Delta{m}^2_{\mathrm{SBL}})
\,.
\label{c-bounds}
\end{equation}
The quantities $a^0_e$ and $a^0_\mu$,
that depend on
$\Delta{m}^2_{\mathrm{SBL}}$,
are given by
\begin{equation}
a^0_\alpha
=
\frac{1}{2}
\left( 1 - \sqrt{ 1 - B_{\alpha\alpha}^0 } \right)
\qquad
(\alpha=e,\mu)
\,,
\label{a0}
\end{equation}
where
$B_{\alpha\alpha}^0$
is the experimental upper bound,
that depends on
$\Delta{m}^2_{\mathrm{SBL}}$,
for the amplitude $B_{\alpha\alpha}$,
obtained from
the exclusion plots of short-baseline
$
\stackrel{\makebox[0pt][l]
{$\hskip-3pt\scriptscriptstyle(-)$}}{\nu_{\alpha}}
$
disappearance experiments (see \cite{BGG-review-98}).
From the 90\% CL
exclusion curves of the Bugey reactor $\bar\nu_e$
disappearance experiment \cite{Bugey-95}
and of the CDHS \cite{CDHS-84}
accelerator $\nu_\mu$ disappearance experiment
it follows that
$ a_e^0 \lesssim 3 \times 10^{-2} $
for
$\Delta{m}^2_{\mathrm{SBL}}$
in the LSND range (\ref{LSND-range})
and
$ a_\mu^0 \lesssim 0.2 $
for
$\Delta m^2_{\mathrm{SBL}} \gtrsim 0.4 \, \mathrm{eV}^2$
(see \cite{BGG-review-98}).

Furthermore,
it is clear that the observed disappearance of
atmospheric $\nu_\mu$'s implies that $\Sigma_\mu$,
that represents the mixing of $\nu_\mu$
with the two massive neutrinos
responsible for atmospheric neutrino oscillations,
cannot be small.
Indeed, 
it has been shown in \cite{BGGS-AB-99}
that the up-down asymmetry of high-energy
$\mu$-like events generated by atmospheric neutrinos
measured in
the Super-Kamiokande experiment
($\mathcal{A}_\mu = 0.311 \pm 0.043 \pm 0.01$
\cite{SK-atm})
implies the bound
\begin{equation}
1 - \Sigma_\mu
\lesssim
0.55
\,.
\label{aSK-old}
\end{equation}
Hence, $\Sigma_\mu$ is large
also for
$\Delta m^2_{\mathrm{SBL}} \lesssim 0.3 \, \mathrm{eV}^2$,
that is below the sensitivity of the short-baseline
$\nu_\mu$ disappearance experiments
performed so far.

The constraints in Eqs.~(\ref{c-bounds}) and (\ref{aSK-old})
show that
$\Sigma_e$ and $1-\Sigma_\mu$ are small
(in both schemes A and B). 
The smallness of $\Sigma_e$ implies that
the oscillations of
$\nu_e$'s
and
$\bar\nu_e$'s
in atmospheric
and long-baseline experiments
are suppressed.
Indeed,
rather strong bounds on the transition probability
$
1 -
P^{\mathrm{LBL}}_{\stackrel{\makebox[0pt][l]
{$\hskip-3pt\scriptscriptstyle(-)$}}{\nu_{e}}
\to\stackrel{\makebox[0pt][l]
{$\hskip-3pt\scriptscriptstyle(-)$}}{\nu_{e}}}
$
of
$
\stackrel{\makebox[0pt][l]
{$\hskip-3pt\scriptscriptstyle(-)$}}{\nu_{e}}
$'s
in long-baseline experiments (LBL)
have been derived in \cite{BGG-bounds-98}.

Using the fact that the oscillations of electron neutrinos
are negligible in atmospheric experiments
and the fact that the Super-Kamiokande up-down asymmetry
has been measured through an average over the
neutrino energy spectrum and propagation distance,
we can derive a lower bound for $\Sigma_\mu$
stronger than the one in Eq.~(\ref{aSK-old}).
We will use the updated \cite{SK-lp99}
value of the Super-Kamiokande
up-down asymmetry,
\begin{equation}
\mathcal{A}_\mu
\equiv
\frac{D-U}{D+U}
=
0.32 \pm 0.04 \pm 0.01
\,,
\label{asy-SK}
\end{equation}
where $D$ and $U$ are the numbers of
down-going and up-going $\mu$-like events,
respectively.
Summing the statistic and systematic errors in quadrature,
we obtain
\begin{equation}
\mathcal{A}_\mu
\geq
0.224
\qquad
\mbox{(99\% CL)}
\,.
\label{asy07}
\end{equation}
Since the asymmetry is due to the disappearance of
$\nu_\mu$'s into
$\nu_\tau$'s or $\nu_s$'s,
we have\footnote{
Here we neglect corrections due to matter effects,
whose calculation would require a complicated fit of the data,
beyond our possibilities.
However,
the matter corrections are expected to be small
because of the average over a wide range of neutrino energies.
}
\begin{equation}
\mathcal{A}_\mu
=
\frac{
\langle
P^{\mathrm{SBL}}_{\stackrel{\makebox[0pt][l]
{$\hskip-3pt\scriptscriptstyle(-)$}}{\nu_{\mu}}
\to\stackrel{\makebox[0pt][l]
{$\hskip-3pt\scriptscriptstyle(-)$}}{\nu_{\mu}}}
\rangle
-
\langle
P^{\mathrm{LBL}}_{\stackrel{\makebox[0pt][l]
{$\hskip-3pt\scriptscriptstyle(-)$}}{\nu_{\mu}}
\to\stackrel{\makebox[0pt][l]
{$\hskip-3pt\scriptscriptstyle(-)$}}{\nu_{\mu}}}
\rangle
}{
\langle
P^{\mathrm{SBL}}_{\stackrel{\makebox[0pt][l]
{$\hskip-3pt\scriptscriptstyle(-)$}}{\nu_{\mu}}
\to\stackrel{\makebox[0pt][l]
{$\hskip-3pt\scriptscriptstyle(-)$}}{\nu_{\mu}}}
\rangle
+
\langle
P^{\mathrm{LBL}}_{\stackrel{\makebox[0pt][l]
{$\hskip-3pt\scriptscriptstyle(-)$}}{\nu_{\mu}}
\to\stackrel{\makebox[0pt][l]
{$\hskip-3pt\scriptscriptstyle(-)$}}{\nu_{\mu}}}
\rangle
}
\,,
\label{asy01}
\end{equation}
where
$
\langle
P^{\mathrm{SBL}}_{\stackrel{\makebox[0pt][l]
{$\hskip-3pt\scriptscriptstyle(-)$}}{\nu_{\mu}}
\to\stackrel{\makebox[0pt][l]
{$\hskip-3pt\scriptscriptstyle(-)$}}{\nu_{\mu}}}
\rangle
$
is the average survival probability of
$
\stackrel{\makebox[0pt][l]
{$\hskip-3pt\scriptscriptstyle(-)$}}{\nu_{\mu}}
$'s
in short-baseline experiments,
that is the same as
the average survival probability of downward-going
atmospheric
$
\stackrel{\makebox[0pt][l]
{$\hskip-3pt\scriptscriptstyle(-)$}}{\nu_{\mu}}
$'s,
and
$
\langle
P^{\mathrm{LBL}}_{\stackrel{\makebox[0pt][l]
{$\hskip-3pt\scriptscriptstyle(-)$}}{\nu_{\mu}}
\to\stackrel{\makebox[0pt][l]
{$\hskip-3pt\scriptscriptstyle(-)$}}{\nu_{\mu}}}
\rangle
$
is the average survival probability of
$
\stackrel{\makebox[0pt][l]
{$\hskip-3pt\scriptscriptstyle(-)$}}{\nu_{\mu}}
$'s
in long-baseline experiments,
that is the same as
the average survival probability of upward-going
atmospheric
$
\stackrel{\makebox[0pt][l]
{$\hskip-3pt\scriptscriptstyle(-)$}}{\nu_{\mu}}
$'s.
From Eqs.~(\ref{LBL08}), (\ref{LBL05}) and
from the definition (\ref{Sigma}),
we have
\begin{eqnarray}
&&
\langle
P^{\mathrm{SBL}}_{\stackrel{\makebox[0pt][l]
{$\hskip-3pt\scriptscriptstyle(-)$}}{\nu_{\mu}}
\to\stackrel{\makebox[0pt][l]
{$\hskip-3pt\scriptscriptstyle(-)$}}{\nu_{\mu}}}
\rangle
=
1 - 2 \Sigma_\mu ( 1 - \Sigma_\mu )
\,,
\label{asy02}
\\
&&
\langle
P^{\mathrm{LBL}}_{\stackrel{\makebox[0pt][l]
{$\hskip-3pt\scriptscriptstyle(-)$}}{\nu_{\mu}}
\to\stackrel{\makebox[0pt][l]
{$\hskip-3pt\scriptscriptstyle(-)$}}{\nu_{\mu}}}
\rangle
=
\langle
P^{\mathrm{SBL}}_{\stackrel{\makebox[0pt][l]
{$\hskip-3pt\scriptscriptstyle(-)$}}{\nu_{\mu}}
\to\stackrel{\makebox[0pt][l]
{$\hskip-3pt\scriptscriptstyle(-)$}}{\nu_{\mu}}}
\rangle
- 2 |U_{\mu\midx{d}}|^2 ( \Sigma_\mu - |U_{\mu\midx{d}}|^2 )
\,.
\label{asy03}
\end{eqnarray}
Taking into account the obvious constraint
$ 0 \leq |U_{\mu\midx{d}}|^2 \leq \Sigma_\mu $,
the long-baseline
average survival probability of
$
\stackrel{\makebox[0pt][l]
{$\hskip-3pt\scriptscriptstyle(-)$}}{\nu_{\mu}}
$'s
is bounded by
\begin{equation}
\langle
P^{\mathrm{SBL}}_{\stackrel{\makebox[0pt][l]
{$\hskip-3pt\scriptscriptstyle(-)$}}{\nu_{\mu}}
\to\stackrel{\makebox[0pt][l]
{$\hskip-3pt\scriptscriptstyle(-)$}}{\nu_{\mu}}}
\rangle
-
\frac{\Sigma_\mu^2}{2}
\leq
\langle
P^{\mathrm{LBL}}_{\stackrel{\makebox[0pt][l]
{$\hskip-3pt\scriptscriptstyle(-)$}}{\nu_{\mu}}
\to\stackrel{\makebox[0pt][l]
{$\hskip-3pt\scriptscriptstyle(-)$}}{\nu_{\mu}}}
\rangle
\leq
\langle
P^{\mathrm{SBL}}_{\stackrel{\makebox[0pt][l]
{$\hskip-3pt\scriptscriptstyle(-)$}}{\nu_{\mu}}
\to\stackrel{\makebox[0pt][l]
{$\hskip-3pt\scriptscriptstyle(-)$}}{\nu_{\mu}}}
\rangle
\,.
\label{asy04}
\end{equation}
Using the lower bound in Eq.~(\ref{asy04}),
for the asymmetry (\ref{asy01}) we have the upper bound
\begin{equation}
\mathcal{A}_\mu
\leq
\frac{\Sigma_\mu^2}{7\Sigma_\mu^2-8\Sigma_\mu+4}
\,.
\label{asy05}
\end{equation}
Inverting this inequality,
we get
\begin{equation}
\Sigma_\mu
\geq
\frac
{4\mathcal{A}_\mu-2\sqrt{\mathcal{A}_\mu(1-3\mathcal{A}_\mu)}}
{7\mathcal{A}_\mu-1}
\,.
\label{asy06}
\end{equation}
This is the lower bound for
$\Sigma_\mu$
given by the asymmetry (\ref{asy01}).
Thus, in the framework of the
four-neutrino schemes under consideration
the asymmetry $\mathcal{A}_\mu$
cannot be larger than 1/3,
a condition that is satisfied by the measured asymmetry (\ref{asy-SK}).
From Eqs.~(\ref{asy07}) and (\ref{asy06}) we obtain
\begin{equation}
1 - \Sigma_\mu
\lesssim
0.38
\equiv
a^{\mathrm{SK}}_\mu
\,.
\label{aSK}
\end{equation}
This is the new upper bound for
$1 - \Sigma_\mu$,
that is more stringent than the one in Eq.~(\ref{aSK-old}).

\begin{figure}
\begin{center}
\mbox{\includegraphics[bb=83 520 305 737,width=0.95\linewidth]{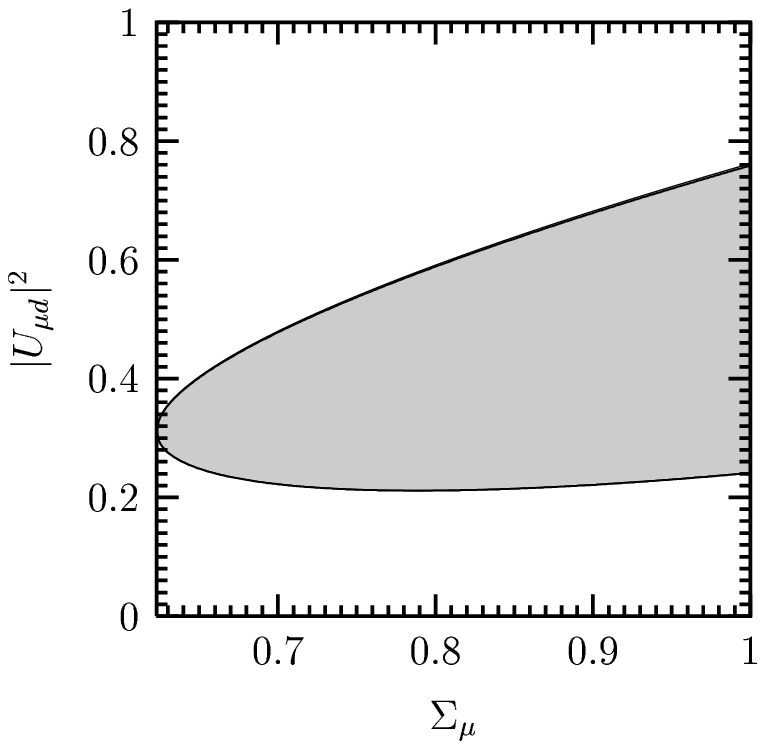}}
\end{center}
\caption{ \label{umu-lim}
Allowed range for $|U_{\mu\midx{d}}|^2$ as a function of
$\Sigma_\mu$
obtained from Eq.~(\ref{asy09})
and
the limit (\ref{asy07}) for the
Super-Kamiokande up-down asymmetry.
}
\end{figure}

Furthermore,
using the expression (\ref{asy03})
for
$
\langle
P^{\mathrm{LBL}}_{\stackrel{\makebox[0pt][l]
{$\hskip-3pt\scriptscriptstyle(-)$}}{\nu_{\mu}}
\to\stackrel{\makebox[0pt][l]
{$\hskip-3pt\scriptscriptstyle(-)$}}{\nu_{\mu}}}
\rangle
$,
we can derive the range of
$|U_{\mu\midx{d}}|^2$
that is allowed by the lower bound
(\ref{asy07})
for $\mathcal{A}_\mu$
(if $|U_{\mu\midx{d}}|^2$
were allowed to vary in the whole range
$[0,\Sigma_\mu]$,
the asymmetry could vanish).
After a straightforward calculation we obtain that
$|U_{\mu\midx{d}}|^2$
is bounded in the range
\begin{equation}
\lambda^{-}_\mu
\leq
|U_{\mu\midx{d}}|^2
\leq
\lambda^{+}_\mu
\,,
\label{asy09}
\end{equation}
with
\begin{equation}
\lambda^{\pm}_\mu
=
\frac{\Sigma_\mu}{2}
\pm
\frac{1}{2}
\,
\sqrt{
\frac
{(1-7\mathcal{A}_\mu)\Sigma_\mu^2-4\mathcal{A}_\mu(1-2\Sigma_\mu)}
{1+\mathcal{A}_\mu}
}
\,.
\label{asy10}
\end{equation}
The allowed range for
$|U_{\mu\midx{d}}|^2$
as a function of $\Sigma_\mu$
obtained from the lower bound (\ref{asy07}) for the asymmetry
is shown by the shadowed area in Fig.~\ref{umu-lim}.
Since $\Sigma_\mu$ can be close to one,
we have the allowed range
\begin{equation}
0.21
\lesssim
|U_{\mu\midx{d}}|^2
\lesssim
0.76
\,.
\label{asy11}
\end{equation}
The same allowed range obviously applies also to
$|U_{\mu\midx{c}}|^2$,
with the constraint (\ref{Sigma})
(for example,
from Fig.~\ref{umu-lim} one can see that
$|U_{\mu\midx{d}}|^2 \simeq 0.21$
is possible only if
$\Sigma_\mu \simeq 0.8$
and the corresponding value of
$|U_{\mu\midx{c}}|^2$
is 0.59,
well within the range (\ref{asy11})).

Up to now we have derived the upper bounds for
$\Sigma_e$ and $1-\Sigma_\mu$
implied by the results of neutrino oscillation experiments.
However,
$\Sigma_e$ and $1-\Sigma_\mu$, albeit small,
cannot vanish,
because non-zero values of both
$\Sigma_e$ and $1-\Sigma_\mu$
are required in order to generate the
$\bar\nu_e\to\bar\nu_\mu$
and
$\nu_e\to\nu_\mu$
transitions observed in the LSND experiment
\cite{LSND}.
Indeed,
we are going to derive lower bounds for
$\Sigma_e$ and $1-\Sigma_\mu$
from the results of the LSND experiment.

Let us consider the amplitude
$A_{{\mu}e}$
of
$\bar\nu_e\to\bar\nu_\mu$
and
$\nu_e\to\nu_\mu$
oscillations in the LSND experiment.
From Eq.~(\ref{SBL-amp}),
using the Cauchy--Schwarz inequality,
we obtain two inequalities:
\begin{eqnarray}
&&
A_{{\mu}e} \leq 4 \Sigma_e \Sigma_\mu
\,,
\label{i1}
\\
&&
A_{{\mu}e} \leq 4 \left( 1 - \Sigma_e \right) \left( 1 - \Sigma_\mu \right)
\,.
\label{i2}
\end{eqnarray}
It is clear that both
$\Sigma_e$ and $1-\Sigma_\mu$
must be bigger than zero and smaller than one in order to have
a non-vanishing amplitude
$A_{{\mu}e}$
in the LSND experiment.

From the inequalities (\ref{i1}) and (\ref{i2})
we obtain
the upper bound
\begin{equation}
1 - \Sigma_\mu \leq
1 - \frac{ A_{{\mu}e}^{\mathrm{min}} }{ 4 a^0_e }
\equiv
a_\mu^{\mathrm{LSND}}
\,,
\label{omcmu-ub}
\end{equation}
already derived in \cite{BGGS-AB-99},
and the lower bounds
\begin{equation}
\Sigma_e \geq \Lambda^{-}_e
\qquad \mbox{and} \qquad
1 - \Sigma_\mu \geq \Lambda^{-}_\mu
\,,
\label{c-lb}
\end{equation}
with
\begin{equation}
\Lambda^{-}_e
=
\Lambda^{-}_\mu
=
\frac{1}{2}
\left( 1 - \sqrt{ 1 - A_{{\mu}e}^{\mathrm{min}} } \right)
\simeq
\frac{ A_{{\mu}e}^{\mathrm{min}} }{ 4 }
\,,
\label{Lambda-}
\end{equation}
where
$A_{{\mu}e}^{\mathrm{min}}$
is the minimum value of the amplitude
$A_{{\mu}e}$
allowed by the results of the LSND experiment
and the approximation is valid for small
$A_{{\mu}e}^{\mathrm{min}}$,
as follows from the LSND data.
Obviously,
the minimum (\ref{Lambda-}) depends on
$\Delta{m}^2_{\mathrm{SBL}}$
trough the dependence from
$\Delta{m}^2_{\mathrm{SBL}}$
of
$A_{{\mu}e}^{\mathrm{min}}$.

Summarizing,
from the results of all neutrino oscillation experiments
we have the bounds
\begin{equation}
\Lambda^{-}_e(\Delta{m}^2_{\mathrm{SBL}})
\leq
\Sigma_e
\leq
\Lambda^{+}_e(\Delta{m}^2_{\mathrm{SBL}})
\,,
\qquad
\Lambda^{-}_\mu(\Delta{m}^2_{\mathrm{SBL}})
\leq
1 - \Sigma_\mu
\leq
\Lambda^{+}_\mu(\Delta{m}^2_{\mathrm{SBL}})
\,,
\label{bounds}
\end{equation}
with
$ \Lambda^{-}_e = \Lambda^{-}_\mu $
given by Eq.(\ref{Lambda-})
and
\begin{equation}
\Lambda^{+}_e
=
a_e^0
\,,
\qquad
\Lambda^{+}_\mu
=
\min\left[
a^0_\mu
\, , \,
a^{\mathrm{SK}}_\mu
\, , \,
a^{\mathrm{LSND}}_\mu
\right]
\,.
\label{Lambda+}
\end{equation}
In Eq.~(\ref{bounds})
we have emphasized that the bounds for
$\Sigma_e$ and $\Sigma_\mu$ depend on the value of
$\Delta{m}^2_{\mathrm{SBL}}$,
that must lie in the LSND range (\ref{LSND-range}).

The dashed line in Fig.~\ref{ce}
represents the lower bound (\ref{c-lb})
for $\Sigma_e$
obtained from the 99\% CL
maximum likelihood allowed region of the LSND
experiment \cite{LSND}.
The solid line in Fig.~\ref{ce}
is the upper bound (\ref{c-bounds})
for $\Sigma_e$
obtained from
the 90\% CL
exclusion curve of the Bugey reactor $\bar\nu_e$
disappearance experiment \cite{Bugey-95}.

The dash-dotted line in Fig.~\ref{ce}
shows the upper bound
for $\Sigma_e$
obtained from the negative result of
the CHOOZ long-baseline reactor $\bar\nu_e$
disappearance experiment \cite{CHOOZ}.
This bound has been obtained taking into account that
the survival probability of $\bar\nu_e$'s in long-baseline experiments
is bounded by
\cite{BGG-bounds-98}
\begin{equation}
P^{\mathrm{LBL}}_{\bar\nu_e\to\bar\nu_e}
\leq
{\Sigma_e}^2 + \left( 1 - \Sigma_e \right)^2
\,.
\label{PeeLBL}
\end{equation}
The CHOOZ experiment has measured a survival probability
averaged over the energy spectrum
\cite{CHOOZ}
\begin{equation}
P^{\mathrm{CHOOZ}}_{\bar\nu_e\to\bar\nu_e}
=
1.010 \pm 0.028 \pm 0.027
=
1.010 \pm 0.039
\,,
\label{PeeCHOOZ}
\end{equation}
where we
added in quadrature
the independent statistical and systematic errors.
Assuming a Gaussian error distribution
and taking into account that the survival probability
is physically limited in the interval $[0,1]$,
we obtain the 90\% CL
lower bound\footnote{
The lower bound (\ref{PeeCHOOZmin})
has been obtained using a Gaussian
distribution normalized in the interval $[0,1]$,
according to the Bayesian prescription
(see \cite{PDG-96,PDG-98,D'Agostini-95}).
The lower bound obtained with the Unified Approach
\cite{Feldman-Cousins-98}
is
$
P^{\mathrm{CHOOZ}}_{\bar\nu_e\to\bar\nu_e}
\geq
0.945
$,
in reasonable agreement with (\ref{PeeCHOOZmin}).
}
\begin{equation}
P^{\mathrm{CHOOZ}}_{\bar\nu_e\to\bar\nu_e}
\geq
0.942
\equiv
P^{\mathrm{CHOOZ,min}}_{\bar\nu_e\to\bar\nu_e}
\,.
\label{PeeCHOOZmin}
\end{equation}
Taking into account the fact that $\Sigma_e$ is small,
from Eqs. (\ref{PeeLBL}) and (\ref{PeeCHOOZmin})
we obtain
\begin{equation}
\Sigma_e
\leq
\frac{1}{2}
\left(
1
-
\sqrt{ 1 - 2 \left( 1 - P^{\mathrm{CHOOZ,min}}_{\bar\nu_e\to\bar\nu_e} \right) }
\right)
=
3.0 \times 10^{-2}
\,.
\label{ce-max-CHOOZ}
\end{equation}
This is the upper bound represented by
the dash-dotted line in Fig.~\ref{ce}.
One can see that the CHOOZ bound
is compatible with the Bugey bound (solid line)
and a more sensitive experiment it is necessary
in order to probe the range of $\Sigma_e$
allowed by the results of the LSND experiment
(shadowed area).

The bounds (\ref{c-bounds}), (\ref{aSK}), (\ref{omcmu-ub}) and (\ref{c-lb})
for $1-\Sigma_\mu$
are illustrated in Fig.~\ref{oms}.
The shadowed area enclosed by the solid line is excluded by the 90\% CL
exclusion curve of the CDHS \cite{CDHS-84}
short-baseline $\nu_\mu$ disappearance experiment
using Eq.~(\ref{c-bounds})
($1 - \Sigma_\mu \leq a^0_\mu$ or $\Sigma_\mu \leq a^0_\mu$
\cite{BGG-AB-96,BGG-review-98}).
The vertically hatched area limited by the dash-dotted line is excluded by
the inequality (\ref{aSK}) obtained
the up-down asymmetry of high-energy
$\mu$-like events generated by atmospheric neutrinos
measured in
the Super-Kamiokande experiment.
The diagonally hatched region is
excluded by the results of the LSND experiment,
taking into account also the
Bugey 90\% CL
exclusion curve.
The region above the dashed line is excluded by the upper bound
(\ref{omcmu-ub})
and the region below the dotted line is excluded by the lower bound
(\ref{c-lb}).
Taking into account all these constraints,
the allowed region for $1-\Sigma_\mu$
is given by the white area in Fig.~\ref{oms}.

\section{Long-baseline disappearance experiments}
\label{Long-baseline disappearance experiments}

From the expression (\ref{LBL04}) with $\alpha=\beta$
(and the corresponding one for antineutrinos)
it follows that
the probability of
$
\stackrel{\makebox[0pt][l]
{$\hskip-3pt\scriptscriptstyle(-)$}}{\nu_{\alpha}}
$
transitions
in other states in long-baseline experiments is bounded by
\cite{BGG-bounds-98}
\begin{equation}
2 \Sigma_\alpha ( 1 - \Sigma_\alpha )
\leq
1
-
P^{\mathrm{LBL}}_{\stackrel{\makebox[0pt][l]
{$\hskip-3pt\scriptscriptstyle(-)$}}{\nu_{\alpha}}
\to\stackrel{\makebox[0pt][l]
{$\hskip-3pt\scriptscriptstyle(-)$}}{\nu_{\alpha}}}
\leq
2 \Sigma_\alpha ( 1 - \frac{1}{2} \, \Sigma_\alpha )
\,.
\label{LBL01}
\end{equation}
It is interesting to notice that the lower bound
coincides with the average probability
of $\nu_\alpha$ transitions
in other states in short-baseline experiments.
Indeed,
from Eqs.~(\ref{Paa-SBL}) and (\ref{SBL-amp})
we have
\begin{equation}
1
-
\langle
P^{\mathrm{SBL}}_{\stackrel{\makebox[0pt][l]
{$\hskip-3pt\scriptscriptstyle(-)$}}{\nu_{\alpha}}
\to\stackrel{\makebox[0pt][l]
{$\hskip-3pt\scriptscriptstyle(-)$}}{\nu_{\alpha}}}
\rangle
=
\frac{1}{2}
\,
B_{\alpha\alpha}
=
2 \Sigma_\alpha ( 1 - \Sigma_\alpha )
\,.
\label{LBL02}
\end{equation}
This is obviously due to the fact
that long-baseline $\nu_\alpha$
disappearance experiments are sensitive
also to short-baseline oscillations due to
$\Delta{m}^2_{\mathrm{SBL}}$,
but since these oscillations are very fast,
only their average is observed in long-baseline experiments.
The surprising fact is that
the interference terms in the long-baseline disappearance probability
are such that the disappearance in long-baseline
experiments cannot be smaller
than the average disappearance in short-baseline experiments.
As we will see later,
the analogous propriety for
$\nu_\alpha\to\nu_\beta$
transitions is not true.

\begin{figure}
\begin{center}
\mbox{\includegraphics[bb=85 520 330 737,width=0.95\linewidth]{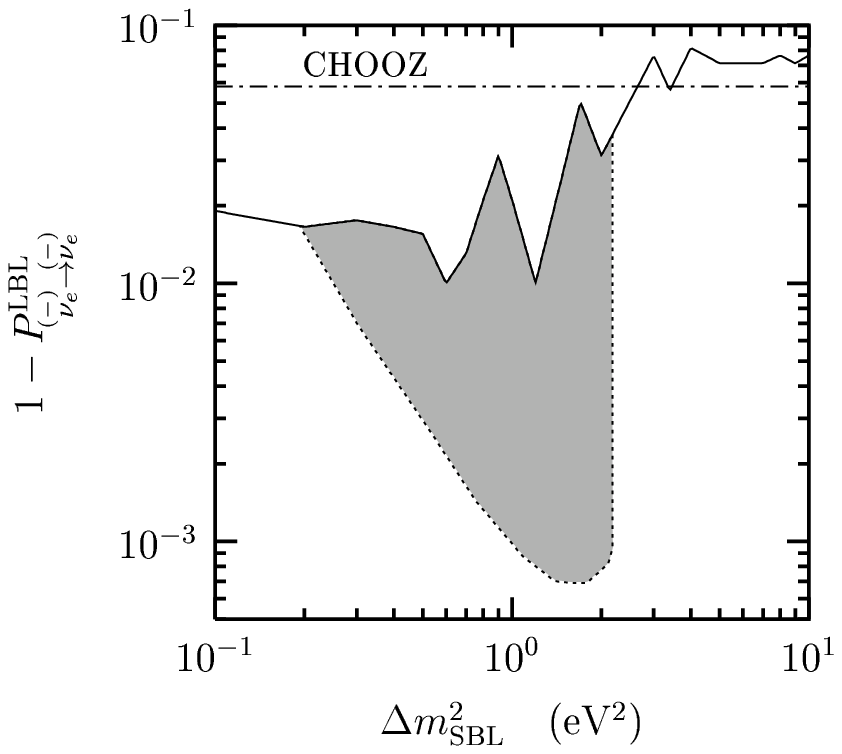}}
\end{center}
\caption{ \label{pee}
Allowed range for
$
1
-
P^{\mathrm{LBL}}_{\stackrel{\protect\makebox[0pt][l]
{$\hskip-3pt\scriptscriptstyle(-)$}}{\nu_{e}}
\to\stackrel{\protect\makebox[0pt][l]
{$\hskip-3pt\scriptscriptstyle(-)$}}{\nu_{e}}}
$
as a function of
$\Delta{m}^2_{\mathrm{SBL}}$
(shadowed region).
The solid line
shows the upper bound in Eq.~(\ref{LBL03})
obtained from the exclusion curve of the Bugey experiment.
The dotted line shows the lower bound in Eq.~(\ref{LBL03})
obtained from the results of the LSND experiment.
The dash-dotted line represents the upper bound
obtained in the CHOOZ experiment.
}
\end{figure}

If we consider $\alpha=e$ in Eq.~(\ref{LBL01}),
taking into account the bound (\ref{bounds}) for $\Sigma_e$,
we have
\begin{equation}
\frac{ A_{{\mu}e}^{\mathrm{min}} }{ 2 }
\simeq
2 \Lambda^{-}_e ( 1 - \Lambda^{-}_e )
\leq
1
-
P^{\mathrm{LBL}}_{\stackrel{\makebox[0pt][l]
{$\hskip-3pt\scriptscriptstyle(-)$}}{\nu_{e}}
\to\stackrel{\makebox[0pt][l]
{$\hskip-3pt\scriptscriptstyle(-)$}}{\nu_{e}}}
\leq
2 \Lambda^{+}_e \left( 1 - \frac{1}{2} \, \Lambda^{+}_e \right)
\simeq
2 a_e^0
\,,
\label{LBL03}
\end{equation}
with the approximations applying, respectively,
for small
$A_{{\mu}e}^{\mathrm{min}}$
and
small
$a_e^0$,
as realized in practice.
Figure~\ref{pee} shows the values of the bounds (\ref{LBL03})
obtained from the existing data.
The solid line in Fig.~\ref{pee}
shows the upper bound in (\ref{LBL03})
obtained from the exclusion curve of the Bugey $\bar\nu_e$
short-baseline disappearance experiment
and the dotted line shows the lower bound in (\ref{LBL03})
obtained from the lower edge of the region
in the $A_{{\mu}e}$--$\Delta{m}^2_{\mathrm{SBL}}$ plane
allowed at 99\% CL by the results of the LSND experiment.
The shadowed region in Fig.~\ref{pee}
is allowed by all data
and one can see that it extends below the CHOOZ bound
(dash-dotted line obtained from Eq.~(\ref{PeeCHOOZmin})).
The lower bound for
$
1
-
P^{\mathrm{LBL}}_{\stackrel{\makebox[0pt][l]
{$\hskip-3pt\scriptscriptstyle(-)$}}{\nu_{\alpha}}
\to\stackrel{\makebox[0pt][l]
{$\hskip-3pt\scriptscriptstyle(-)$}}{\nu_{\alpha}}}
$
lies above $7 \times 10^{-4}$,
a limit that will be very difficult to reach.
However,
if $\Delta{m}^2_{\mathrm{SBL}}$
lies in the lower part of the LSND-allowed region,
the lower bound
$
1
-
P^{\mathrm{LBL}}_{\stackrel{\makebox[0pt][l]
{$\hskip-3pt\scriptscriptstyle(-)$}}{\nu_{\alpha}}
\to\stackrel{\makebox[0pt][l]
{$\hskip-3pt\scriptscriptstyle(-)$}}{\nu_{\alpha}}}
$
is substantially higher
and may be reachable in the near future
(see \cite{Mikaelyan-99}).

Considering the disappearance of $\nu_\mu$'s
in long-baseline experiments,
it is clear that,
since $\Sigma_\mu$ is large,
the corresponding probability can vary
almost from zero to one.
Indeed,
using the limits (\ref{bounds}) for $\Sigma_\mu$
in Eq.(\ref{LBL01}) we obtain
\begin{equation}
\frac{ A_{{\mu}e}^{\mathrm{min}} }{ 2 }
\simeq
2 \Lambda^{-}_\mu ( 1 - \Lambda^{-}_\mu )
\leq
1
-
P^{\mathrm{LBL}}_{\stackrel{\makebox[0pt][l]
{$\hskip-3pt\scriptscriptstyle(-)$}}{\nu_{\mu}}
\to\stackrel{\makebox[0pt][l]
{$\hskip-3pt\scriptscriptstyle(-)$}}{\nu_{\mu}}}
\leq
\left( 1 - \Lambda^{-}_\mu \right)
\left( 1 + \frac{1}{2} \, \Lambda^{-}_\mu \right)
\simeq
1 - \frac{ A_{{\mu}e}^{\mathrm{min}} }{ 8 }
\,.
\label{LBL051}
\end{equation}

The expression (\ref{LBL01})
gives the bounds for the transition probability
$
1
-
P^{\mathrm{LBL}}_{\stackrel{\makebox[0pt][l]
{$\hskip-3pt\scriptscriptstyle(-)$}}{\nu_{\alpha}}
\to\stackrel{\makebox[0pt][l]
{$\hskip-3pt\scriptscriptstyle(-)$}}{\nu_{\alpha}}}
$,
that depends on the neutrino energy.
However,
if a long baseline experiment
is sensitive to values of $\Delta{m}^2$
smaller than the actual value of $\Delta{m}^2_{\mathrm{atm}}$,
its transition probability is given by
the average (\ref{LBL05}).
For $\alpha=\beta$ in Eq.~(\ref{LBL05}),
taking into account the
definition of $\Sigma_\alpha$ in Eq.~(\ref{Sigma})
and the obvious constraints
$|U_{\alpha\midx{c}}|^2, |U_{\alpha\midx{d}}|^2 \leq \Sigma_\alpha$,
one can derive the bounds
\begin{equation}
2 \Sigma_\alpha ( 1 - \Sigma_\alpha )
\leq
1
-
\langle
P^{\mathrm{LBL}}_{\stackrel{\makebox[0pt][l]
{$\hskip-3pt\scriptscriptstyle(-)$}}{\nu_{\alpha}}
\to\stackrel{\makebox[0pt][l]
{$\hskip-3pt\scriptscriptstyle(-)$}}{\nu_{\alpha}}}
\rangle
\leq
2 \Sigma_\alpha \left( 1 - \frac{3}{4} \, \Sigma_\alpha \right)
\,.
\label{LBL06}
\end{equation}
Therefore,
only the upper bound for the average transition probability is
different from the corresponding one for the transition probability,
but the difference in practice is negligible for
$\alpha=e$,
because $\Sigma_e$ is very small.
Hence,
the bounds in Fig.~\ref{pee}
apply also to
$
1
-
\langle
P^{\mathrm{LBL}}_{\stackrel{\makebox[0pt][l]
{$\hskip-3pt\scriptscriptstyle(-)$}}{\nu_{e}}
\to\stackrel{\makebox[0pt][l]
{$\hskip-3pt\scriptscriptstyle(-)$}}{\nu_{e}}}
\rangle
$.

For the average transition probability
of $\nu_\mu$'s into other states,
a lower bound that is stronger than the lower bound
implied by Eq.~(\ref{LBL06})
(that coincides with the one in Eq.~(\ref{LBL051}))
is given by the asymmetry (\ref{asy01}),
from which we get
\begin{equation}
\langle
P^{\mathrm{LBL}}_{\stackrel{\makebox[0pt][l]
{$\hskip-3pt\scriptscriptstyle(-)$}}{\nu_{\mu}}
\to\stackrel{\makebox[0pt][l]
{$\hskip-3pt\scriptscriptstyle(-)$}}{\nu_{\mu}}}
\rangle
=
\frac{1-\mathcal{A}_\mu}{1+\mathcal{A}_\mu}
\,
\langle
P^{\mathrm{SBL}}_{\stackrel{\makebox[0pt][l]
{$\hskip-3pt\scriptscriptstyle(-)$}}{\nu_{\mu}}
\to\stackrel{\makebox[0pt][l]
{$\hskip-3pt\scriptscriptstyle(-)$}}{\nu_{\mu}}}
\rangle
\,.
\label{asy21}
\end{equation}
Taking into account the limit (\ref{asy07})
for $\mathcal{A}_\mu$
and approximating the upper bound
$
\langle
P^{\mathrm{SBL}}_{\stackrel{\makebox[0pt][l]
{$\hskip-3pt\scriptscriptstyle(-)$}}{\nu_{\mu}}
\to\stackrel{\makebox[0pt][l]
{$\hskip-3pt\scriptscriptstyle(-)$}}{\nu_{\mu}}}
\rangle
\leq
1
-
2 \Lambda^{-}_\mu ( 1 - \Lambda^{-}_\mu )
$
with one,
we obtain
\begin{equation}
1
-
\langle
P^{\mathrm{LBL}}_{\stackrel{\makebox[0pt][l]
{$\hskip-3pt\scriptscriptstyle(-)$}}{\nu_{\mu}}
\to\stackrel{\makebox[0pt][l]
{$\hskip-3pt\scriptscriptstyle(-)$}}{\nu_{\mu}}}
\rangle
\gtrsim
0.37
\,.
\label{asy22}
\end{equation}
This is the lower bound for the
average transition probability of
$
\stackrel{\makebox[0pt][l]
{$\hskip-3pt\scriptscriptstyle(-)$}}{\nu_{\mu}}
$'s
into other states in order to
accommodate the up-down asymmetry observed
in the Super-Kamiokande experiment.
We have also an upper bound for this transition
probability,
that is given by Eq.~(\ref{LBL06}):
\begin{equation}
1
-
\langle
P^{\mathrm{LBL}}_{\stackrel{\makebox[0pt][l]
{$\hskip-3pt\scriptscriptstyle(-)$}}{\nu_{\mu}}
\to\stackrel{\makebox[0pt][l]
{$\hskip-3pt\scriptscriptstyle(-)$}}{\nu_{\mu}}}
\rangle
\leq
\left\{
\begin{array}{ll}
\left( 1 - \Lambda^{+}_\mu \right)
\left( 1 + 3 \Lambda^{+}_\mu \right) / 2
&
\quad \mbox{for} \quad
\Lambda^{+}_\mu < 1/3
\,,
\\
2/3
&
\quad \mbox{for} \quad
\Lambda^{+}_\mu \geq 1/3
\,,
\end{array}
\right.
\label{LBL061}
\end{equation}
where we have taken into account the fact that,
from Eqs.~(\ref{aSK}) and (\ref{Lambda+}),
$\Lambda^{+}_\mu$
can be as large as about 0.38.
It is interesting to notice that there is the possibility
that
$
1
-
\langle
P^{\mathrm{LBL}}_{\stackrel{\makebox[0pt][l]
{$\hskip-3pt\scriptscriptstyle(-)$}}{\nu_{\mu}}
\to\stackrel{\makebox[0pt][l]
{$\hskip-3pt\scriptscriptstyle(-)$}}{\nu_{\mu}}}
\rangle
$
is larger that $1/2$.
Since in the case of two-neutrino mixing
the average transition probability
of any flavor neutrino into other states
is smaller than 1/2
(equal in the case of maximal mixing),
the measurement in long-baseline experiments of
an average transition probability
of $\nu_\mu$'s into other states
larger than 1/2
would be an evidence of the existence of a neutrino
mass-squared difference larger than the atmospheric one
and more than two-neutrino mixing.
Equation (\ref{LBL061})
shows that this is possible in the framework
of four-neutrino mixing,
given the existing experimental data.

\section{Long-baseline appearance experiments}
\label{Long-baseline appearance experiments}

Taking into account the definition (\ref{SBL-amp})
for $A_{\alpha\beta}$,
from Eq.~(\ref{LBL04})
it follows that the probability
of
$
\stackrel{\makebox[0pt][l]
{$\hskip-3pt\scriptscriptstyle(-)$}}{\nu_{\alpha}}
\to\stackrel{\makebox[0pt][l]
{$\hskip-3pt\scriptscriptstyle(-)$}}{\nu_{\beta}}
$
transitions
with $\alpha\neq\beta$
in long-baseline experiments is bounded by
\cite{BGG-bounds-98}
\begin{equation}
\frac{1}{4} \, A_{\alpha\beta}
\leq
P^{\mathrm{LBL}}_{\stackrel{\makebox[0pt][l]
{$\hskip-3pt\scriptscriptstyle(-)$}}{\nu_{\alpha}}
\to\stackrel{\makebox[0pt][l]
{$\hskip-3pt\scriptscriptstyle(-)$}}{\nu_{\beta}}}
\leq
\frac{1}{4} \, A_{\alpha\beta}
+
\Sigma_\alpha \Sigma_\beta
\,,
\label{LBL07}
\end{equation}
and by the unitarity upper limits
\begin{equation}
P^{\mathrm{LBL}}_{\stackrel{\makebox[0pt][l]
{$\hskip-3pt\scriptscriptstyle(-)$}}{\nu_{\alpha}}
\to\stackrel{\makebox[0pt][l]
{$\hskip-3pt\scriptscriptstyle(-)$}}{\nu_{\beta}}}
\leq
1
-
P^{\mathrm{LBL}}_{\stackrel{\makebox[0pt][l]
{$\hskip-3pt\scriptscriptstyle(-)$}}{\nu_{\alpha}}
\to\stackrel{\makebox[0pt][l]
{$\hskip-3pt\scriptscriptstyle(-)$}}{\nu_{\alpha}}}
\,,
\qquad
P^{\mathrm{LBL}}_{\stackrel{\makebox[0pt][l]
{$\hskip-3pt\scriptscriptstyle(-)$}}{\nu_{\alpha}}
\to\stackrel{\makebox[0pt][l]
{$\hskip-3pt\scriptscriptstyle(-)$}}{\nu_{\beta}}}
\leq
1
-
P^{\mathrm{LBL}}_{\stackrel{\makebox[0pt][l]
{$\hskip-3pt\scriptscriptstyle(-)$}}{\nu_{\beta}}
\to\stackrel{\makebox[0pt][l]
{$\hskip-3pt\scriptscriptstyle(-)$}}{\nu_{\beta}}}
\,.
\label{LBL07-uni}
\end{equation}
The corresponding allowed range for
$
P^{\mathrm{LBL}}_{\stackrel{\makebox[0pt][l]
{$\hskip-3pt\scriptscriptstyle(-)$}}{\nu_{\mu}}
\to\stackrel{\makebox[0pt][l]
{$\hskip-3pt\scriptscriptstyle(-)$}}{\nu_{e}}}
$
obtained from the experimental data
has been presented in Ref.~\cite{BGG-bounds-98}.
The other constrained channels are
$\nu_\alpha\to\nu_\beta$
with
$\alpha=e,\mu$
and
$\beta=\tau,s$,
for which we have only the upper limits
\begin{equation}
P^{\mathrm{LBL}}_{\stackrel{\makebox[0pt][l]
{$\hskip-3pt\scriptscriptstyle(-)$}}{\nu_{\alpha}}
\to\stackrel{\makebox[0pt][l]
{$\hskip-3pt\scriptscriptstyle(-)$}}{\nu_{\beta}}}
\leq
1
-
P^{\mathrm{LBL}}_{\stackrel{\makebox[0pt][l]
{$\hskip-3pt\scriptscriptstyle(-)$}}{\nu_{\alpha}}
\to\stackrel{\makebox[0pt][l]
{$\hskip-3pt\scriptscriptstyle(-)$}}{\nu_{\alpha}}}
\qquad
(
\alpha=e,\mu
\,
\quad
\beta=\tau,s
)
\,,
\label{LBL071}
\end{equation}
due to the conservation of probability,
with the upper limits
for
$
1
-
P^{\mathrm{LBL}}_{\stackrel{\makebox[0pt][l]
{$\hskip-3pt\scriptscriptstyle(-)$}}{\nu_{e}}
\to\stackrel{\makebox[0pt][l]
{$\hskip-3pt\scriptscriptstyle(-)$}}{\nu_{e}}}
$
and
$
1
-
P^{\mathrm{LBL}}_{\stackrel{\makebox[0pt][l]
{$\hskip-3pt\scriptscriptstyle(-)$}}{\nu_{\mu}}
\to\stackrel{\makebox[0pt][l]
{$\hskip-3pt\scriptscriptstyle(-)$}}{\nu_{\mu}}}
$
given in Eqs.~(\ref{LBL03}) and (\ref{LBL051}),
respectively.

Confronting the interval (\ref{LBL07}) with the expression for the
average probability of
$
\stackrel{\makebox[0pt][l]
{$\hskip-3pt\scriptscriptstyle(-)$}}{\nu_{\alpha}}
\to\stackrel{\makebox[0pt][l]
{$\hskip-3pt\scriptscriptstyle(-)$}}{\nu_{\beta}}
$
transitions
in short-baseline experiments given in Eq.~(\ref{LBL08}),
one can see that,
as anticipated in the discussion after Eq.~(\ref{LBL02}),
the probability of long-baseline
$
\stackrel{\makebox[0pt][l]
{$\hskip-3pt\scriptscriptstyle(-)$}}{\nu_{\alpha}}
\to\stackrel{\makebox[0pt][l]
{$\hskip-3pt\scriptscriptstyle(-)$}}{\nu_{\beta}}
$
transitions
can be smaller than the average probability of short-baseline
$
\stackrel{\makebox[0pt][l]
{$\hskip-3pt\scriptscriptstyle(-)$}}{\nu_{\alpha}}
\to\stackrel{\makebox[0pt][l]
{$\hskip-3pt\scriptscriptstyle(-)$}}{\nu_{\beta}}
$
transitions.
This is due to interference effects produced by
$\Delta{m}^2_{\mathrm{atm}}$.
Hence we have
\begin{equation}
P^{\mathrm{LBL}}_{\stackrel{\makebox[0pt][l]
{$\hskip-3pt\scriptscriptstyle(-)$}}{\nu_{\alpha}}
\to\stackrel{\makebox[0pt][l]
{$\hskip-3pt\scriptscriptstyle(-)$}}{\nu_{\beta}}}
<
\langle
P^{\mathrm{SBL}}_{\stackrel{\makebox[0pt][l]
{$\hskip-3pt\scriptscriptstyle(-)$}}{\nu_{\alpha}}
\to\stackrel{\makebox[0pt][l]
{$\hskip-3pt\scriptscriptstyle(-)$}}{\nu_{\beta}}}
\rangle
\qquad
\mbox{(possible)}
\,.
\label{LBL09}
\end{equation}
For a fixed source-detector distance in a long-baseline experiment,
the realization of the possible inequality
(\ref{LBL09})
depends on the value of the neutrino energy.
Therefore,
using a detector with sufficient energy resolution
(or with a sufficiently narrow-band beam
and varying the neutrino energy)
it is possible to check experimentally
if the inequality (\ref{LBL09}) can be realized.
The inequality (\ref{LBL09})
is interesting because it can be realized only if there are
at least two mass-squared differences,
$\Delta{m}^2_{\mathrm{SBL}}$
and
$\Delta{m}^2_{\mathrm{atm}}$.
Indeed,
it is impossible in the case of two-neutrino mixing
that implies,
with an obvious notation,
\begin{eqnarray}
&&
P^{\mathrm{LBL},2}_{\stackrel{\makebox[0pt][l]
{$\hskip-3pt\scriptscriptstyle(-)$}}{\nu_{\alpha}}
\to\stackrel{\makebox[0pt][l]
{$\hskip-3pt\scriptscriptstyle(-)$}}{\nu_{\beta}}}
=
\langle
P^{\mathrm{SBL},2}_{\stackrel{\makebox[0pt][l]
{$\hskip-3pt\scriptscriptstyle(-)$}}{\nu_{\alpha}}
\to\stackrel{\makebox[0pt][l]
{$\hskip-3pt\scriptscriptstyle(-)$}}{\nu_{\beta}}}
\rangle
\qquad
\mbox{only $\Delta{m}^2_{\mathrm{SBL}}$}
\,,
\label{LBL11}
\\
&&
P^{\mathrm{LBL},2}_{\stackrel{\makebox[0pt][l]
{$\hskip-3pt\scriptscriptstyle(-)$}}{\nu_{\alpha}}
\to\stackrel{\makebox[0pt][l]
{$\hskip-3pt\scriptscriptstyle(-)$}}{\nu_{\beta}}}
\geq
\langle
P^{\mathrm{SBL},2}_{\stackrel{\makebox[0pt][l]
{$\hskip-3pt\scriptscriptstyle(-)$}}{\nu_{\alpha}}
\to\stackrel{\makebox[0pt][l]
{$\hskip-3pt\scriptscriptstyle(-)$}}{\nu_{\beta}}}
\rangle
=
0
\qquad
\mbox{only $\Delta{m}^2_{\mathrm{atm}}$}
\,.
\label{LBL12}
\end{eqnarray}
However,
the possibility of the inequality
(\ref{LBL09})
is not unique to the four-neutrino schemes,
and it can be realized, for example,
in three-neutrino schemes
with
$\Delta{m}^2_{\mathrm{31}}=\Delta{m}^2_{\mathrm{SBL}}$
and
$\Delta{m}^2_{\mathrm{21}}=\Delta{m}^2_{\mathrm{atm}}$.

The experimental check of
the inequality (\ref{LBL09})
could provide some information on the relative phases
of the elements of the mixing matrix.
Indeed,
confronting Eqs.~(\ref{LBL04}) and (\ref{LBL08}),
taking into account of the definition (\ref{SBL-amp})
for $A_{\alpha\beta}$,
one can see that the inequality (\ref{LBL09}) for neutrinos
is realized if
\begin{equation}
\mathrm{Re}\!\left[
U_{\alpha\midx{c}}
U_{\beta\midx{c}}^{*}
U_{\alpha\midx{d}}^{*}
U_{\beta\midx{d}}
\exp\!\left(
- i
\frac{ \Delta{m}^{2}_{\mathrm{atm}}  L }{ 2  E }
\right)
\right]
<
\mathrm{Re}\!\left[
U_{\alpha\midx{c}}
U_{\beta\midx{c}}^{*}
U_{\alpha\midx{d}}^{*}
U_{\beta\midx{d}}
\right]
\,.
\label{LBL13}
\end{equation}
The product
$
U_{\alpha\midx{c}}
U_{\beta\midx{c}}^{*}
U_{\alpha\midx{d}}^{*}
U_{\beta\midx{d}}
$
is invariant under rephasing of the mixing matrix,
\textit{i.e.}
the transformations
$U_{\alpha k} \to e^{i\gamma_\alpha} U_{\alpha k}$
and the transformations
$U_{\alpha k} \to e^{i\gamma_k} U_{\alpha k}$.
Using these transformations
we can consider, without loss of generality,
$U_{\alpha\midx{c}}=|U_{\alpha\midx{c}}|$,
$U_{\beta\midx{c}}=|U_{\beta\midx{c}}|$,
$U_{\alpha\midx{d}}=|U_{\alpha\midx{d}}|$,
$U_{\beta\midx{d}}=|U_{\beta\midx{d}}|e^{i\xi}$
and write the inequality (\ref{LBL13}) as
\begin{equation}
\cos( \xi - \varphi )
<
\cos\xi
\,,
\label{LBL14}
\end{equation}
where
$\varphi \equiv \Delta{m}^2_{\mathrm{atm}}L/2E$.
If we confine
$\xi$
in the interval $[0,2\pi)$,
the inequality (\ref{LBL14})
is not realized for any value of $\varphi$
only if
\begin{equation}
\xi = \pi
\,,
\label{LBL15}
\end{equation}
and it is realized for all values of $\varphi\neq0$
only if
\begin{equation}
\xi = 0
\,.
\label{LBL16}
\end{equation}
Since
the two extreme cases (\ref{LBL15}) and (\ref{LBL16})
correspond to the absence of CP-violating phases
in the $\alpha,\beta$-$1,2$
sector of the mixing matrix,
it is clear that the observation of the inequality
(\ref{LBL09})
for some, but not all values of $\varphi$
would be an evidence in favor of CP violation
in the lepton sector due to neutrino mixing.
This fact can also be seen by noting that
if CP is conserved
all the elements of the mixing matrix are real
and
there are two possibilities:
\begin{eqnarray}
&&
1)
\quad
\frac{1}{4} \, A_{\alpha\beta}
=
\left|
U_{\alpha\midx{c}}
U_{\beta\midx{c}}
+
U_{\alpha\midx{d}}
U_{\beta\midx{d}}
\right|^2
=
\left(
|U_{\alpha\midx{c}}U_{\beta\midx{c}}|
+
|U_{\alpha\midx{d}}U_{\beta\midx{d}}|
\right)^2
\quad
\mbox{if}
\quad
U_{\alpha\midx{c}}
U_{\beta\midx{c}}
U_{\alpha\midx{d}}
U_{\beta\midx{d}}
>
0
\,,
\label{LBL17}
\\
&&
2)
\quad
\frac{1}{4} \, A_{\alpha\beta}
=
\left|
U_{\alpha\midx{c}}
U_{\beta\midx{c}}
+
U_{\alpha\midx{d}}
U_{\beta\midx{d}}
\right|^2
=
\left(
|U_{\alpha\midx{c}}U_{\beta\midx{c}}|
-
|U_{\alpha\midx{d}}U_{\beta\midx{d}}|
\right)^2
\quad
\mbox{if}
\quad
U_{\alpha\midx{c}}
U_{\beta\midx{c}}
U_{\alpha\midx{d}}
U_{\beta\midx{d}}
<
0
\,.
\label{LBL18}
\end{eqnarray}
In the first case we have
\begin{equation}
1)
\qquad
\left|
U_{\alpha\midx{c}}
U_{\beta\midx{c}}
+
U_{\alpha\midx{d}}
U_{\beta\midx{d}}
e^{-i\varphi}
\right|^2
\leq
\frac{1}{4} \, A_{\alpha\beta}
\,,
\label{LBL19}
\end{equation}
and
\begin{equation}
1)
\qquad
P^{\mathrm{LBL}}_{\stackrel{\makebox[0pt][l]
{$\hskip-3pt\scriptscriptstyle(-)$}}{\nu_{\alpha}}
\to\stackrel{\makebox[0pt][l]
{$\hskip-3pt\scriptscriptstyle(-)$}}{\nu_{\beta}}}
\leq
\langle
P^{\mathrm{SBL}}_{\stackrel{\makebox[0pt][l]
{$\hskip-3pt\scriptscriptstyle(-)$}}{\nu_{\alpha}}
\to\stackrel{\makebox[0pt][l]
{$\hskip-3pt\scriptscriptstyle(-)$}}{\nu_{\beta}}}
\rangle
\qquad
\mbox{always}
\,.
\label{LBL20}
\end{equation}
In the second case we have
\begin{equation}
2)
\qquad
\left|
U_{\alpha\midx{c}}
U_{\beta\midx{c}}
+
U_{\alpha\midx{d}}
U_{\beta\midx{d}}
e^{-i\varphi}
\right|^2
\geq
\frac{1}{4} \, A_{\alpha\beta}
\,,
\label{LBL21}
\end{equation}
and
\begin{equation}
2)
\qquad
P^{\mathrm{LBL}}_{\stackrel{\makebox[0pt][l]
{$\hskip-3pt\scriptscriptstyle(-)$}}{\nu_{\alpha}}
\to\stackrel{\makebox[0pt][l]
{$\hskip-3pt\scriptscriptstyle(-)$}}{\nu_{\beta}}}
\geq
\langle
P^{\mathrm{SBL}}_{\stackrel{\makebox[0pt][l]
{$\hskip-3pt\scriptscriptstyle(-)$}}{\nu_{\alpha}}
\to\stackrel{\makebox[0pt][l]
{$\hskip-3pt\scriptscriptstyle(-)$}}{\nu_{\beta}}}
\rangle
\qquad
\mbox{always}
\,.
\label{LBL22}
\end{equation}
The inequality (\ref{LBL22}) has the same origin as the inequality
(\ref{LBL12}) in the case of two-neutrino mixing.

If
$\Delta{m}^2_{\mathrm{atm}}$
is known and
the the detector has a sufficient energy resolution
(or with a sufficiently narrow-band beam
and varying the neutrino energy),
it is even possible to measure the CP-violating phase $\xi$
by looking at the variation of the
transition probability as a function of $\varphi$.
For example, we have
\begin{eqnarray}
&&
P^{\mathrm{LBL}}_{\nu_\alpha\to\nu_\beta}
(\varphi=\pi)
-
P^{\mathrm{LBL}}_{\nu_\alpha\to\nu_\beta}
(\varphi=0)
=
- 4
|U_{\alpha\midx{c}} U_{\beta\midx{c}} U_{\alpha\midx{d}} U_{\beta\midx{d}}|
\cos\xi
\,,
\label{LBL23}
\\
&&
P^{\mathrm{LBL}}_{\nu_\alpha\to\nu_\beta}
(\varphi=\frac{3}{2}\pi)
-
P^{\mathrm{LBL}}_{\nu_\alpha\to\nu_\beta}
(\varphi=\frac{1}{2}\pi)
=
- 4
|U_{\alpha\midx{c}} U_{\beta\midx{c}} U_{\alpha\midx{d}} U_{\beta\midx{d}}|
\sin\xi
\,.
\label{LBL24}
\end{eqnarray}
The ratio of Eqs.~(\ref{LBL24}) and (\ref{LBL23})
gives $\tan\xi$
independently from the value of
$
|U_{\alpha\midx{c}} U_{\beta\midx{c}} U_{\alpha\midx{d}} U_{\beta\midx{d}}|
$.

Let us finally consider the average transition probabilities
that are measured in long-baseline experiments
that are sensitive to values of $\Delta{m}^2$
much smaller than the actual value of
$\Delta{m}^2_{\mathrm{atm}}$.
Considering
$
\stackrel{\makebox[0pt][l]
{$\hskip-3pt\scriptscriptstyle(-)$}}{\nu_{\mu}}
\to\stackrel{\makebox[0pt][l]
{$\hskip-3pt\scriptscriptstyle(-)$}}{\nu_{e}}
$
transitions,
from Eq.~(\ref{LBL05}) we have
\begin{equation}
\langle
P^{\mathrm{LBL}}_{\stackrel{\makebox[0pt][l]
{$\hskip-3pt\scriptscriptstyle(-)$}}{\nu_{\mu}}
\to\stackrel{\makebox[0pt][l]
{$\hskip-3pt\scriptscriptstyle(-)$}}{\nu_{e}}}
\rangle
=
\frac{1}{4}
\,
A_{{\mu}e}
+
|U_{\mu\midx{c}}|^2
|U_{e\midx{c}}|^2
+
|U_{\mu\midx{d}}|^2
|U_{e\midx{d}}|^2
\,.
\label{LBL31}
\end{equation}
Taking into account the bounds (\ref{bounds})
for $\Sigma_e$ and $\Sigma_\mu$,
the allowed range (\ref{asy09}) for
$|U_{\mu\midx{d}}|^2 = \Sigma_\mu - |U_{\mu\midx{c}}|^2$
and
the constraint
$0 \leq |U_{e\midx{d}}|^2 \leq \Sigma_e$
for
$|U_{e\midx{d}}|^2 = \Sigma_e - |U_{e\midx{c}}|^2$,
one can find that the average transition probability (\ref{LBL31})
is bounded by
\begin{equation}
\frac{1}{4}
\,
A_{{\mu}e}
+
\lambda^{-}_\mu(\mathrm{min})
\Lambda^{-}_e
\leq
\langle
P^{\mathrm{LBL}}_{\stackrel{\makebox[0pt][l]
{$\hskip-3pt\scriptscriptstyle(-)$}}{\nu_{\mu}}
\to\stackrel{\makebox[0pt][l]
{$\hskip-3pt\scriptscriptstyle(-)$}}{\nu_{e}}}
\rangle
\leq
\frac{1}{4}
\,
A_{{\mu}e}
+
\lambda^{+}_\mu(\mathrm{max})
\Lambda^{+}_e
\,.
\label{LBL32}
\end{equation}
The lower bound is reached for
$|U_{e\midx{d}}|^2=0$
and
$|U_{\mu\midx{d}}|^2=\Sigma_\mu-\lambda^{-}_\mu(\mathrm{min})$
or
$|U_{e\midx{d}}|^2=\Sigma_e$
and
$|U_{\mu\midx{d}}|^2=\lambda^{-}_\mu(\mathrm{min})$.
The quantity
$\lambda^{-}_\mu(\mathrm{min})$
is the minimum value of
$\lambda^{-}_\mu$
in the allowed range for $\Sigma_\mu$.
For the values of $\Delta{m}^2_{\mathrm{SBL}}$
in which $\Sigma_\mu$ is limited by the Super-Kamiokande
asymmetry constraint (\ref{aSK}),
the minimum of $\lambda^{-}_\mu$,
shown in Fig.~\ref{umu-lim},
occurs for
\begin{equation}
\Sigma_\mu
=
\frac{
8\mathcal{A}_\mu
-
\sqrt{2(1-3\mathcal{A}_\mu)(1+\mathcal{A}_\mu)}
}{
2(7\mathcal{A}_\mu-1)
}
\,.
\label{LBL33}
\end{equation}
The upper bound in Eq.(\ref{LBL32}) is reached for
$|U_{e\midx{d}}|^2=0$
and
$|U_{\mu\midx{d}}|^2=\Sigma_\mu-\lambda^{+}_\mu(\mathrm{max})$
or
$|U_{e\midx{d}}|^2=\Sigma_e$
and
$|U_{\mu\midx{d}}|^2=\lambda^{+}_\mu(\mathrm{max})$.
The quantity
$\lambda^{+}_\mu(\mathrm{max})$
is the maximum value of
$\lambda^{+}_\mu$
in the allowed range for $\Sigma_\mu$,
that,
as shown in Fig.~\ref{umu-lim},
occurs always for the maximum allowed value of $\Sigma_\mu$,
$\Sigma_\mu^{\mathrm{max}} = 1 - \Lambda^{-}_\mu$.

\begin{figure}
\begin{center}
\mbox{\includegraphics[bb=85 520 330 737,width=0.95\linewidth]{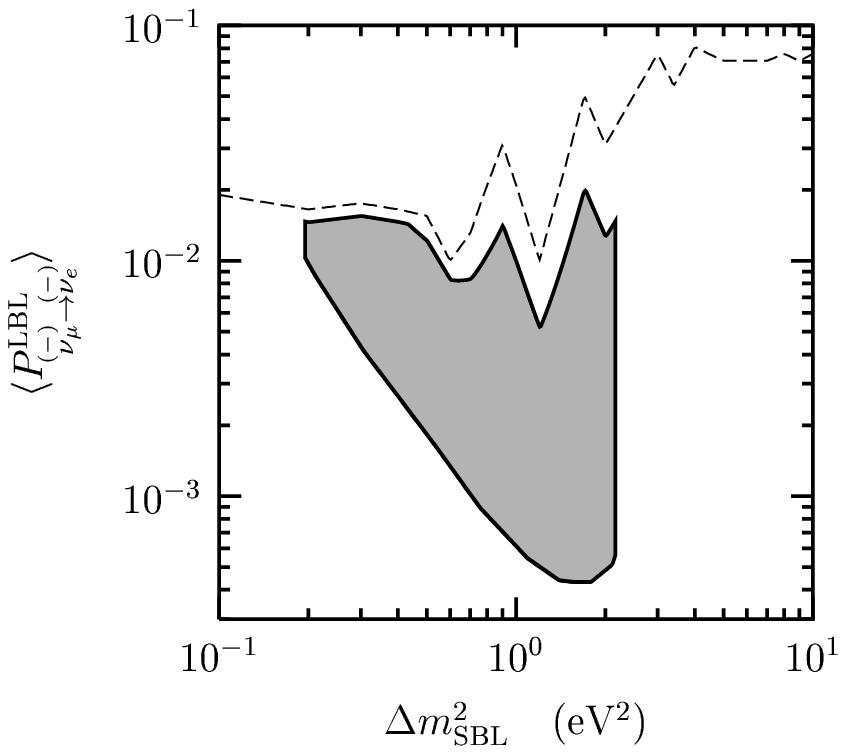}}
\end{center}
\caption{ \label{pme}
Allowed range (\ref{LBL32}) for
$
\langle
P^{\mathrm{LBL}}_{\stackrel{\protect\makebox[0pt][l]
{$\hskip-3pt\scriptscriptstyle(-)$}}{\nu_{\mu}}
\to\stackrel{\protect\makebox[0pt][l]
{$\hskip-3pt\scriptscriptstyle(-)$}}{\nu_{e}}}
\rangle
$
as a function of
$\Delta{m}^2_{\mathrm{SBL}}$
(shadowed region).
The dashed curve represents the unitarity constraint
$
\langle
P^{\mathrm{LBL}}_{\stackrel{\protect\makebox[0pt][l]
{$\hskip-3pt\scriptscriptstyle(-)$}}{\nu_{\mu}}
\to\stackrel{\protect\makebox[0pt][l]
{$\hskip-3pt\scriptscriptstyle(-)$}}{\nu_{e}}}
\rangle
\leq
1
-
\langle
P^{\mathrm{LBL}}_{\stackrel{\protect\makebox[0pt][l]
{$\hskip-3pt\scriptscriptstyle(-)$}}{\nu_{e}}
\to\stackrel{\protect\makebox[0pt][l]
{$\hskip-3pt\scriptscriptstyle(-)$}}{\nu_{e}}}
\rangle
$.
}
\end{figure}

Figure~\ref{pme} shows the allowed range (\ref{LBL32})
as a function of
$\Delta{m}^2_{\mathrm{SBL}}$
(shadowed region).
The dashed line in Fig.~\ref{pme} represents
the unitarity upper bound
$
\langle
P^{\mathrm{LBL}}_{\stackrel{\makebox[0pt][l]
{$\hskip-3pt\scriptscriptstyle(-)$}}{\nu_{\mu}}
\to\stackrel{\makebox[0pt][l]
{$\hskip-3pt\scriptscriptstyle(-)$}}{\nu_{e}}}
\rangle
\leq
1
-
\langle
P^{\mathrm{LBL}}_{\stackrel{\makebox[0pt][l]
{$\hskip-3pt\scriptscriptstyle(-)$}}{\nu_{e}}
\to\stackrel{\makebox[0pt][l]
{$\hskip-3pt\scriptscriptstyle(-)$}}{\nu_{e}}}
\rangle
$,
with the upper bound for
$
1
-
\langle
P^{\mathrm{LBL}}_{\stackrel{\makebox[0pt][l]
{$\hskip-3pt\scriptscriptstyle(-)$}}{\nu_{e}}
\to\stackrel{\makebox[0pt][l]
{$\hskip-3pt\scriptscriptstyle(-)$}}{\nu_{e}}}
\rangle
$
given by Eq.~(\ref{LBL06}).
Since $\Lambda^{+}_e=a_e^0$
is small,
the dashed line in Fig.~\ref{pme}
practically coincides with the solid line in Fig.~\ref{pee}.

The other constrained channels are
$\nu_\alpha\to\nu_\beta$
with
$\alpha=e,\mu$
and
$\beta=\tau,s$,
for which
$
\langle
P^{\mathrm{LBL}}_{\stackrel{\makebox[0pt][l]
{$\hskip-3pt\scriptscriptstyle(-)$}}{\nu_{\alpha}}
\to\stackrel{\makebox[0pt][l]
{$\hskip-3pt\scriptscriptstyle(-)$}}{\nu_{\beta}}}
\rangle
$
has only the unitarity upper limits
given by the average of Eq.~(\ref{LBL071}).

\section{Conclusions}
\label{Conclusions}

\begin{figure}
\begin{center}
\mbox{\includegraphics[bb=95 520 330 737,width=0.95\linewidth]{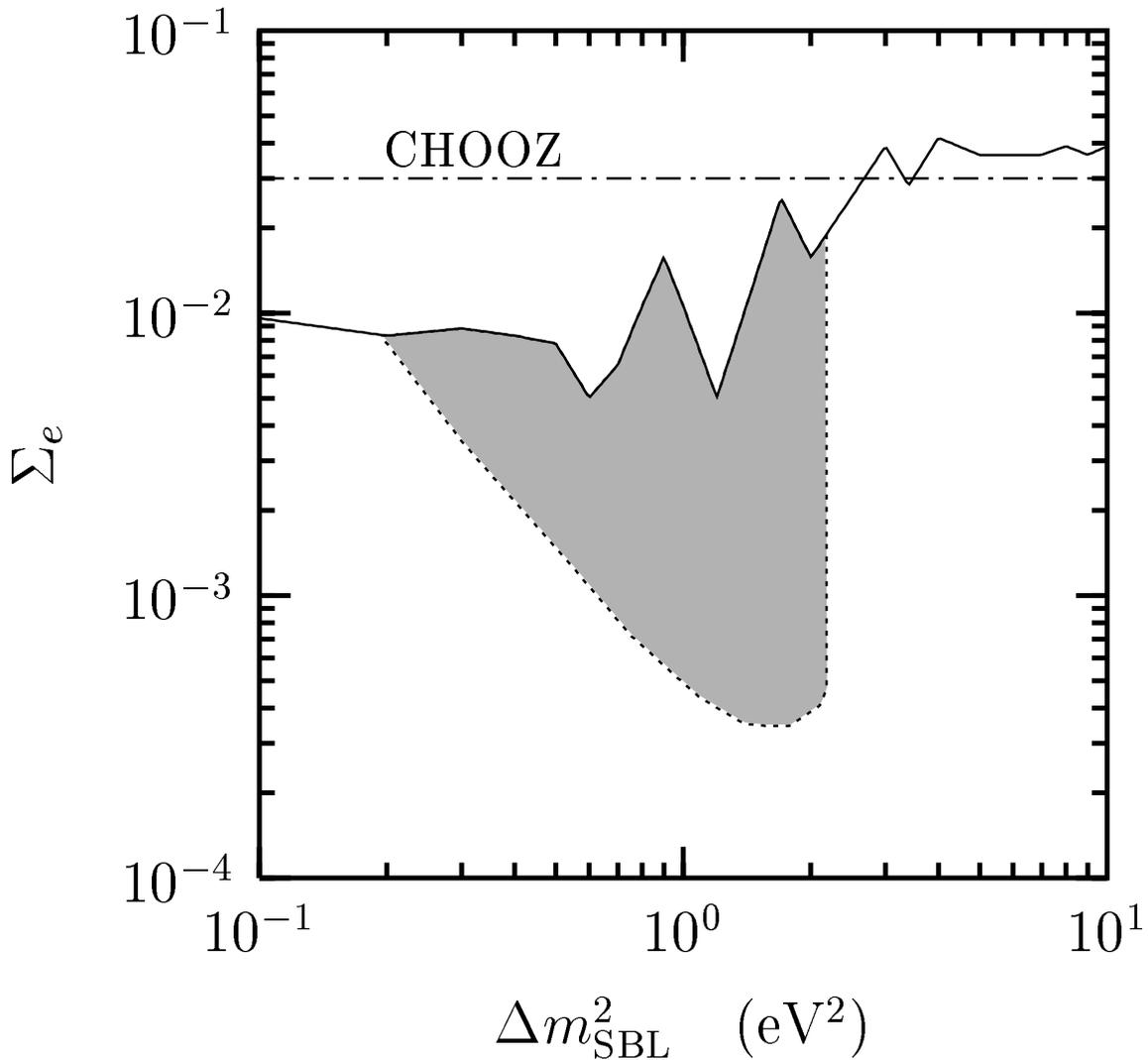}}
\end{center}
\caption{ \label{ce}
Allowed range for $\Sigma_e$ as a function of
$\Delta{m}^2_{\mathrm{SBL}}$
(shadowed region).
The solid line 
is the upper bound (\ref{c-bounds})
obtained from
the exclusion curve of the Bugey experiment.
The dashed line
represents the lower bound (\ref{c-lb})
obtained from the allowed region of the LSND
experiment.
The dash-dotted line
shows the upper bound (\ref{ce-max-CHOOZ})
obtained from the negative result of
the CHOOZ experiment.
}
\end{figure}

\begin{figure}
\begin{center}
\mbox{\includegraphics[bb=95 520 330 737,width=0.95\linewidth]{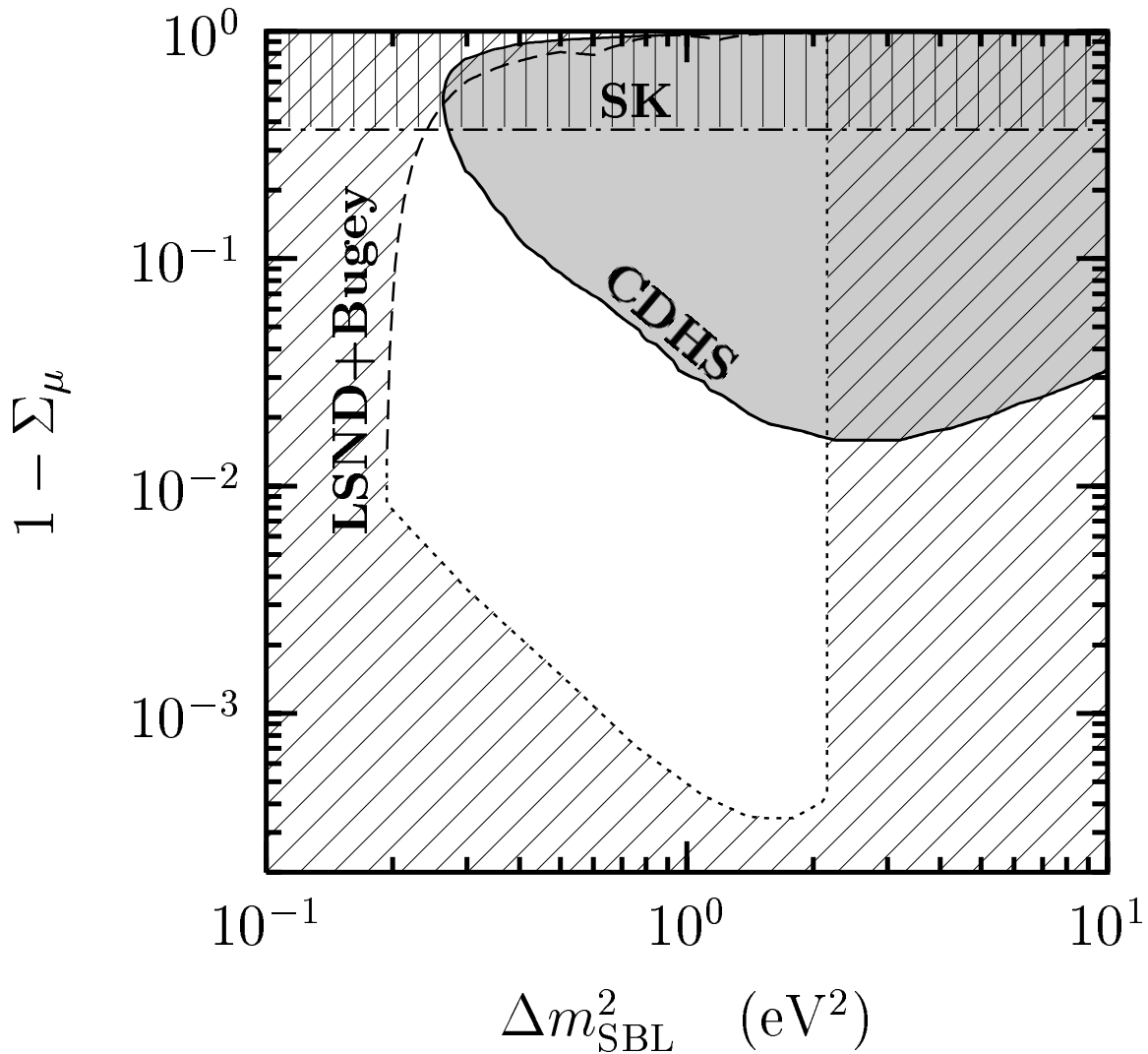}}
\end{center}
\caption{ \label{oms}
Allowed range for $\Sigma_\mu$ as a function of
$\Delta{m}^2_{\mathrm{SBL}}$
(white region).
The shadowed area enclosed by the solid line is excluded by the 90\% CL
exclusion curve of the CDHS experiment.
The vertically hatched area limited by the dash-dotted line is excluded by
the inequality (\ref{aSK}) obtained from the
Super-Kamiokande up-down asymmetry.
The region above the dashed line is excluded by the upper bound
(\ref{omcmu-ub})
and the region below the dotted line is excluded by the lower bound
(\ref{c-lb}).
}
\end{figure}

We have presented the constraints on the neutrino mixing parameters
obtained from the results of neutrino oscillation experiments
in the framework of the two four-neutrino schemes (\ref{AB})
that are compatible with all neutrino oscillation data.
We have shown that the parameters
$\Sigma_e$ and $1-\Sigma_\mu$
defined in Eq.~(\ref{Sigma})
are small,
but do not vanish.
Their allowed ranges are presented in Figs.~\ref{ce} and \ref{oms}.
We have also derived the bounds for the mixing
of $\nu_\mu$
with the two massive neutrinos
whose mass-squared difference is responsible for
atmospheric neutrino oscillations (Eq.~(\ref{asy09})).
In Sections
\ref{Long-baseline disappearance experiments}
and
\ref{Long-baseline appearance experiments}
we presented the bounds for the
oscillation probabilities
in long-baseline experiments
(some of which have been derived already in Ref.~\cite{BGG-bounds-98}).
We have also shown that
a comparison of the probability measured in long-baseline
appearance experiments
with the corresponding average probability measured
in short-baseline experiments
may give information on the relative phases
of the relevant elements of the mixing matrix
and, in particular,
may provide an evidence for the existence of CP violation
in the lepton sector.
This method may be more convenient than the
traditional method based on the comparison of the
oscillation probabilities of neutrinos and antineutrinos
if only a neutrino or antineutrino beam is available.

In this paper we have considered long-baseline experiments
neglecting matter effects,
that are rather complicated in the framework
of four-neutrino mixing
and will be discussed in detail in another article
\cite{DGKK-preparation}.
We think that
it is important to discuss long-baseline oscillations
in vacuum for at least two reasons.
First,
because vacuum oscillations
can be treated analytically
and all the effects can be understood in a clear way.
The matter effects,
that often must be calculated numerically,
can be appreciated as corrections
(sometimes large)
to the oscillations in vacuum.
Second,
in order to measure accurately the properties of neutrinos
without the uncertainty and complications
caused by matter effects,
it may be possible,
in some indeterminate future,
to perform a long-baseline experiment in vacuum.
For example,
it may be possible to produce a neutrino beam
in a orbital station
(existing space stations orbit at an altitude
between 300 and 400 km)
and shoot it
to some detector(s) on the Earth.
This experiment would allow to measure accurately the properties of neutrinos
when the beam does not cross the Earth
and,
with this knowledge,
to perform an accurate tomography of the Earth
using the matter effects
when the beam crosses the Earth.
At present such an experiment is fanciful,
but we know that many experiments
have been performed up to now
that were unthinkable or believed to be impossible in the past.


\newpage

\end{document}